\documentclass[journal, onecolumn]{IEEEtran}

\usepackage{lineno,hyperref}

\modulolinenumbers[5]

\usepackage[level]{datetime}
\usepackage{lineno}
\usepackage{graphicx}  
\graphicspath{{figures/}}  
\usepackage{subfigure}  
\usepackage{amsmath}
\usepackage{multirow}   
\usepackage[noend]{algpseudocode}    
\usepackage{algorithmicx,algorithm}  
\usepackage{color}

\usepackage{bm}                       
\usepackage{amssymb}                  
\usepackage{amsbsy} 

\usepackage{amsmath}

\ifCLASSINFOpdf
\else
\fi
\begin{document}
%
\title{Efficient Compressed Sensing Based Image Coding by Using Gray Transformation}
%
%
%

\author{Bo~Zhang,~\IEEEmembership{}
        Di~Xiao,~\IEEEmembership {Member,~IEEE}, Lan Wang, Sen~Bai,~\IEEEmembership{} and Lei Yang~\IEEEmembership{}
\thanks{This work was supported in part by the National Key R$\&$D Program of China (Grant No. 2020YFB1805400), in part by the National Natural Science Foundation of China (Grant No. 62072063),  in part by the Chongqing Research Program of Basic Research and Frontier Technology (Grant No. cstc2017jcyjBX0008), in part by the Scientific and Technological Research Program of Chongqing Municipal Education Commission (Grant No. KJZD-K201801901), in part by the Chongqing Postgraduate Education Reform Project (Grant No. yjg183018) and in part by the Chongqing University Postgraduate Education Reform Project (Grant No. cquyjg18219).}
\thanks{B. Zhang and L. Yang are with the Communication NCO Academy, Army Engineering University, Chongqing 400035, China (e-mail: zhangboswjtu@163.com; yangleicqcc@126.com).}
\thanks{D. Xiao is with the Key Laboratory of Dependable Service Computing in Cyber Physical Society of Ministry of Education, College of Computer
Science, Chongqing University, Chongqing 400044, China (e-mail: xiaodi\_cqu@hotmail.com).}
\thanks{L. Wang is with the School of Mathematics and Information Engineering, Chongqing University of Education, Chongqing 400147, China (e-mail: wanglan@cque.edu.cn).}
\thanks{S. Bai is with Chongqing Institute of Engineering, Chongqing 400056, China (e-mail: baisencq@126.com).}
\thanks{\textit{Corresponding author: Di Xiao}.}}

%
%

\markboth{Efficient Compressed Sensing Based Image Coding by Using Gray Transformation}%
{Shell \MakeLowercase{\textit{et al.}}: Bare Demo of IEEEtran.cls for IEEE Journals}
%



\maketitle

\begin{abstract}
In recent years, compressed sensing (CS) based image coding has become a hot topic in image processing field. However, since the bit depth required for encoding each CS sample is too large, the compression performance of this paradigm is unattractive. To address this issue, a novel CS-based image coding system by using gray transformation is proposed. In the proposed system, we use a gray transformation to preprocess the original image firstly and then use CS to sample the transformed image. Since gray transformation makes the probability distribution of CS samples centralized, the bit depth required for encoding each CS sample is reduced significantly. Consequently, the proposed system can considerably improve the compression performance of CS-based image coding. Simulation results show that the proposed system outperforms the traditional one without using gray transformation in terms of compression performance.
\end{abstract}


%
\IEEEpeerreviewmaketitle

\section{Introduction}
%
%
%
%
\IEEEPARstart{I}{n} the last decade, compressed sensing (CS) [1], [2] has attracted a lot of interest in signal processing field, which states that a sparse signal or a compressible signal can be efficiently captured and recovered by using a small number of linear samples. Compared with the classical Shannon-Nyquist sampling theorem, CS can recover the original signal by using far fewer samples.

In recent years, image coding by using CS has become a hot topic in image processing field [3]-[15]. The block diagram of the traditional CS-based image coding system is shown in Fig. 1. Firstly, the original image is sampled by using CS. Then, the CS samples are quantized and entropy coded into bits. According to classical quantization theory, for a given quantization step size, the quantization distortion depends on the probability distribution of source symbols. Generally, the more centralized the source symbols distribute, the smaller the quantization distortion is. Unfortunately, the distribution of CS samples is usually decentralized [3], which means that the quantization distortion is very large if the bit depth allocated for each CS sample is too small. One way to solve this problem is to increase the bit depth for each CS sample. For example, when Bernoulli random matrix is used, it has been shown that at least 10 bits are needed to encode each CS sample [4]. However, this method will increase the bit rate significantly, which makes CS-based image coding unattractive in terms of the compression performance.

There are a number of studies on improving the compression performance of CS-based image coding, largely through an optimization of the sampling process (e.g., [5]-[8]), the quantization process (e.g., [9]-[12]), or the reconstruction process (e.g., [13]-[15]). However, for these methods, the distribution of CS samples is still decentralized. If we can find a strategy to make the distribution of CS samples more centralized, the compression performance can be furtherly improved.
\begin{figure*}
  \centering
  \includegraphics[width=0.65 \linewidth]{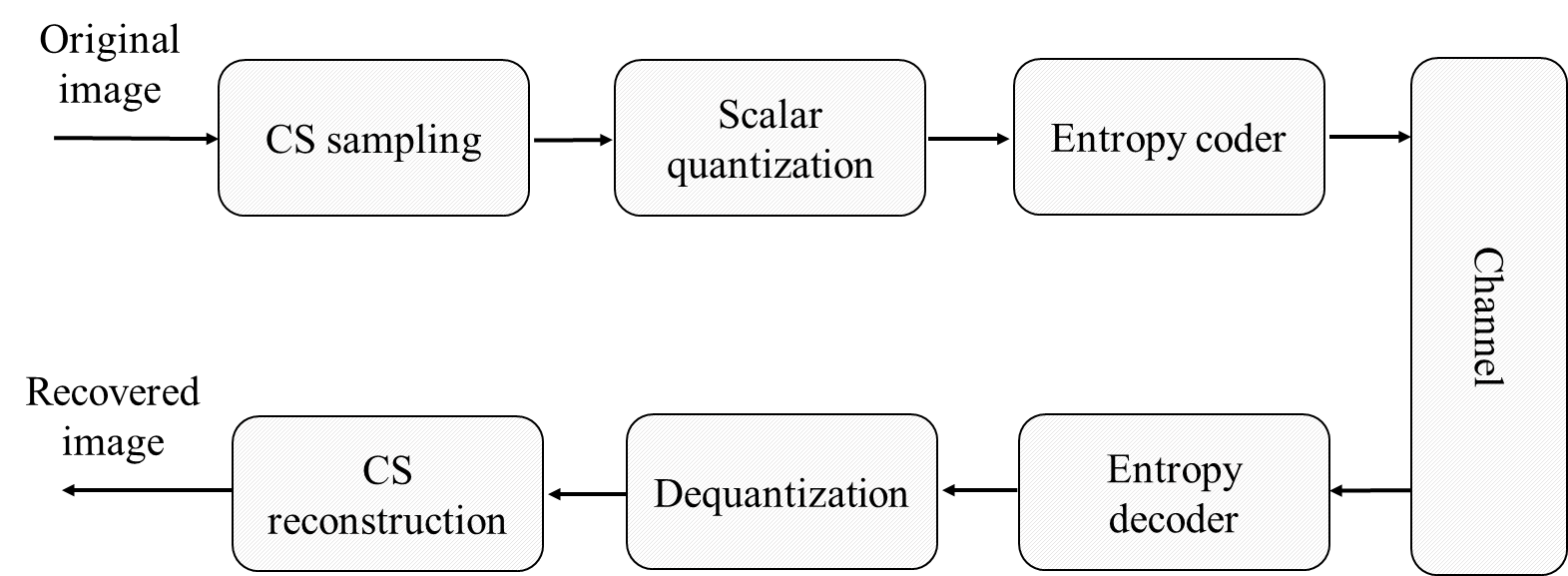}
  \caption{The traditional CS-based image coding system.}
  \end{figure*}

In this letter, we propose a novel CS-based image coding system by using gray transformation. In the proposed system, we use a gray transformation to preprocess the original image firstly and then use CS to sample the transformed image. Since gray transformation can decrease the variance of CS samples significantly, i.e., this transformation makes the probability distribution of CS samples more centralized, the quantization distortion of CS samples is reduced significantly. Consequently, the proposed method can considerably improve the compression performance of CS-based image coding.

We organize the rest of this letter as follows. The theory of CS is given in Section II. The motivation of this work is presented in Section III. The proposed image coding system is described in detail in Section IV. Some simulation results are given in Section V. Finally, we concludes this work in Section VI.

\section{CS Backgrounds}

Consider a 1D signal $\textbf{\emph{x}}\in R^{N}$, which can be represented as
$$
\textbf{\emph{x}}=\bm{\varPsi}\textbf{\emph{s}},\eqno(1)
$$ 	
where $ \bm{\varPsi}\in R^{{N}\times {N}} $ is orthogonal basis matrix and $\textbf{\emph{s}}\in R^{N}$ is a vector which is composed by coefficients in the $ \bm{\varPsi} $  domain. If most of the coefficients of $\textbf{\emph{s}} $  are zero, we say the signal $\textbf{\emph{x}} $ is a sparse signal.

Suppose that  $ \bm{\varPhi} \in R ^ {M\times N} \left(M\ll N\right) $  is a measurement matrix, then the CS samples of $\textbf{\emph{x}} $  can be obtained by using a linear projection operation, i. e.,
$$
\textbf{\emph{y}}=\bm{\varPhi}\textbf{\emph{x}},\eqno(2)
$$
where $ \textbf{\emph{y}}\in R^{M}$  is the CS measurement vector of $ \textbf{\emph{x}} $. The sampling rate is defined as $ SR=\frac{M}{N}$.
	
According to CS theory, if the measurement matrix satisfies restricted isometry property (RIP), $ \textbf{\emph{y}} $ can be used to recover the sparse signal accurately. It has been demonstrated that Gaussian random matrices satisfy RIP with high probability.

At the decoder side, the image can be reconstructed by solving the following  $ l_1 $-norm minimization problem

$$
\hat{\textbf{\emph{x}}}=\operatorname*{argmin} \limits_{{\textbf{\emph{x}}}}
\parallel\bm{\varPsi}^{\rm T}\textbf{\emph{x}}\parallel_1
 \ {\rm s}. \ {\rm t}. \quad \textbf{\emph{y}}=
\bm{\varPhi} \textbf{\emph{x}}.\eqno(3)
$$

In practice, the above optimization problem can be solved effectively by using any CS reconstruction algorithm [16]-[18].

\section{The motivation of the proposed method}

Before describing our method in detail, the motivation of the proposed method is presented.

It should be note that CS only focus on reducing the number of samples. In order to realize a real compression, the CS samples needs to be quantized and entropy coded into bits.

When quantization and entropy coding are taken into consideration, the bit rate for CS-based image coding is given as
$$
R_x=\frac{M}{N}R_y,\eqno(4)
$$
where $ R_y $  is the rate of the quantizer which is lower bounded by the entropy of the quantized CS samples.

In this letter, the measurement matrix is selected to be Gaussian random matrix. When Gaussian random matrix is used, the CS samples obey Gaussian distribution [10]. According to [10], the distortion for the CS samples is

$$
D\left(R_y\right)=\frac{1}{6}\pi \delta ^2 2^{-2R_y},\eqno(5)
$$
where $ \delta $  is the variance of the CS samples.

For a given $ R_y $, it can be seen from Eq. (5) that the distortion of CS samples is dominated by the variance of CS samples. If we can find a method to reduce the variance, the distortion can be reduced significantly. This motivates us to develop a strategy to reduce the variance of CS samples, thus improving the compression performance of CS-based image coding.

\section{The proposed method}

In this letter, we propose a novel CS-based image coding system by using gray transformation. The overall architecture of the proposed system is shown in Fig. 2. At the encoder side, we preprocess the original image by gray transformation. After gray transformation, we can get the transformed image. Then, the transformed image is sampled by using CS. Thirdly, the CS measurement vector is quantized into integer index by scalar quantization (SQ). Lastly, the integer index is furtherly compressed into bits by using lossless Huffman coding. At the decoder side, by using inverse Huffman coding and de-quantization operation, we can recover the de-quantized CS measurement vector of the transformed image firstly. Then, we can locally recover the de-quantized CS measurement vector of the original image. Finally, the original image can be reconstructed by using any CS reconstruction algorithm. Since gray transformation can reduce the variance of CS samples significantly, the proposed system can considerably improve the compression performance of CS-based image coding. The detailed steps of the proposed method will be elaborated below.

\begin{figure*}
  \centering
  \includegraphics[width= 0.85 \linewidth]{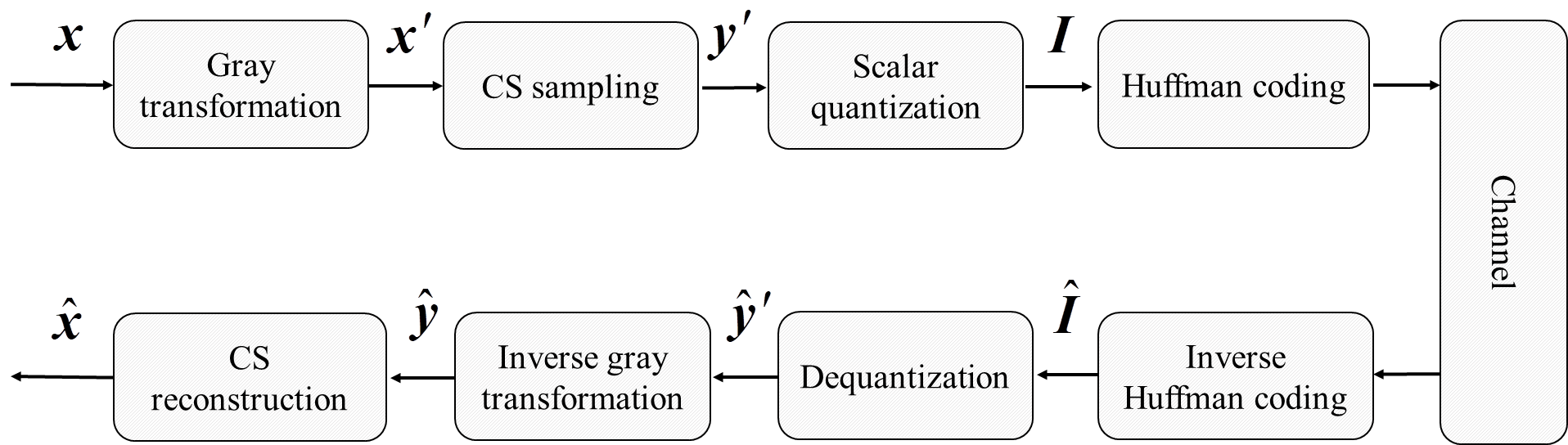}
  \caption{The proposed CS-based image coding system.}
  \end{figure*}

\subsection{Image coding}

Suppose that $ \textbf{\emph{x}}\in R^{N} $ is the original image, which is obtained by stacking the 2D image $ \textbf{\emph{X}}\in R^{\sqrt{N} \times \sqrt{N}} $ in a column-by-column manner. In order to improve the compression performance, we preprocess the 2D image by using gray transformation firstly and then sample the transformed image. The detailed steps of image coding includes four steps.

\textbf{Step 1}: Use a gray transformation to preprocess the original image, which can be formulated as
$$
\textbf{\emph{x}}'=\textbf{\emph{x}}-\textbf{\emph{a}},\eqno(6)
$$
where $ \textbf{\emph{a}} = \left[a, a, ..., a\right] \in R^{N} $  is a constant vector and $ \textbf{\emph{x}}' \in R^{N} $  is the transformed image.

\textbf{Step 2}: the transformed image is sampled by using CS

$$
\textbf{\emph{y}}'=\bm{\varPhi}\textbf{\emph{x}}' = \bm{\varPhi}\textbf{\emph{x}}-\bm{\varPhi}\textbf{\emph{a}}=\textbf{\emph{y}}-\textbf{\emph{y}}_a,\eqno(7)
$$
where $\textbf{\emph{y}}'\in R^{M}$,  $\textbf{\emph{y}}\in R^{M} $ and $\textbf{\emph{y}}_a\in R^{M} $  denote the CS measurement vectors of  $\textbf{\emph{x}}' $,  $\textbf{\emph{x}} $ and $\textbf{\emph{a}} $, respectively.

Obviously, $\textbf{\emph{y}}_a $ does not contain any information because it can be regenerated by $\textbf{\emph{a}} $ in the decoder side. Therefore, if the receiver obtains $\textbf{\emph{y}}' $ and the constant $ a $, he can recover the CS measurement vector $\textbf{\emph{y}} $. However, by appropriately selecting the constant $ a $, $\textbf{\emph{y}}' $ will has smaller variance than  $\textbf{\emph{y}} $, thus reducing the distortion significantly. The detailed steps to select an appropriate constant $ a $ will be discussed in Subsection IV-C.

\textbf{Step 3}: The CS measurement vector is quantized into integer index by using scalar quantization (SQ). The integer quantization index of  $\textbf{\emph{y}}' $ can be obtained by
$$
\textbf{\emph{I}}= \operatorname*{Q}\left(\textbf{\emph{y}}'\right), \eqno(8)
$$
 where $ \operatorname*{Q}\left(\cdot\right) $   is referred as the forward quantization stage.

\textbf{Step 4}: We encode the quantization index and the constant $ a $ into bits by using Huffman coding.

\subsection{Image reconstruction}

We will describe the detailed steps of image reconstruction in this subsection. It includes three steps.

\textbf{Step 1}: Recover the CS measurement vector of the transformed image by using inverse Huffman coding and dequantization operation. The de-quantized value of  $ \textbf{\emph{y}}' $ can be obtained by
$$
\hat {\textbf{\emph{y}}}' = \operatorname*{Q^{-1}} \left( \operatorname*{Q}\left(\textbf{\emph{y}}'\right)\right), \eqno(9)
$$
where  $ \operatorname*{Q^{-1}}\left(\cdot\right) $ refers as the inverse quantization stage and  $ \hat{\textbf{\emph{y}}}'\in R^{M} $  is the de-quantized CS measurement vector of  $ \textbf{\emph{x}}' $.

\textbf{Step 2}: The CS measurement vector of the original image can be locally recovered by inverse gray transformation operation

$$
\hat{\textbf{\emph{y}}}= \hat{\textbf{\emph{y}}}' + \textbf{\emph{y}}_a,\eqno(10)
$$
where $ \hat{\textbf{\emph{y}}}\in R^{M} $  is the de-quantized CS measurement vector of  $ \textbf{\emph{x}} $ and $ \textbf{\emph{y}}_a=\bm{\varPhi}\textbf{\emph{a}} $ can be locally regenerated by using the constant $ a $.

\textbf{Step 3}: Use the CS measurement vector to recover the original image
$$
\hat  {\textbf{\emph{x}}} = \operatorname*{CS^{-1}}\left(\hat  {\textbf{\emph{y}}}\right), \eqno(11)
$$
where $ \hat  {\textbf{\emph{x}}}\in R^{N} $  is the reconstructed image and   $ \operatorname*{CS^{-1}} \left(\cdot\right)$ denotes an CS reconstruction function.

\subsection{Optimal constant vector selection}

A novel CS-based image coding system by using gray transformation is proposed in this letter. By appropriately selecting the constant vector,  $ \textbf{\emph{y}}' $  will has smaller variance than $ \textbf{\emph{y}} $, thus reducing the distortion significantly. However, for a given image, how to select an proper constant vector is not discussed yet. In this subsection, we will study this problem.

\textbf{\emph {Theorem 1}} Suppose the entries of $ \bm{\varPhi} $  are i.i.d. Gaussian variables with mean zero and variance  $ 1/M $, then the CS measurement vector  obeys Gauss distribution. The mean and the variance of $ \textbf{\emph{y}}' $  are $ \operatorname*{E}\left(\textbf{\emph{y}}'_i\right) =0 $  and $ \operatorname*{Var}\left(\textbf{\emph{y}}'_i\right)= \frac{1}{M} \sum\limits_{j=1}^N \left( \textbf{\emph{x}}_j -a \right)^2 $, respectively.

\textbf{\emph {Proof}}: Obviously, the $i $-th entries of  $ \textbf{\emph{y}}' $ is  $ \textbf{\emph{y}}'_i = \sum\limits_{j=1}^N \bm{\varPhi} _{ij} \left( \textbf{\emph{x}}_j -a \right)$ . Since  $ \textbf{\emph{y}}'_i $ is a sum of Gaussian variables,  $ \textbf{\emph{y}}'_i $  is still a Gaussian variable. Therefore,  $ \textbf{\emph{y}}' $ obeys Gauss distribution. The mean value and the variance of $ \textbf{\emph{y}}' $  can be written as
 $$
\operatorname*{E}\left(\textbf{\emph{y}}'_i\right) =
\operatorname*{E}\left( \sum\limits_{j=1}^N \bm{\varPhi} _{ij} \left( \textbf{\emph{x}}_j -a \right)\right)
$$
$$
=\operatorname*{E}\left( \bm{\varPhi} _{ij}\right)\sum\limits_{j=1}^N \left( \textbf{\emph{x}}_j -a \right)=0
, \eqno(12)
$$
and
$$
\operatorname*{Var}\left(\textbf{\emph{y}}'_i\right) =
\operatorname*{Var}\left( \sum\limits_{j=1}^N \bm{\varPhi} _{ij} \left( \textbf{\emph{x}}_j -a \right)\right) \quad \quad
$$
$$
 \quad \quad  =\sum\limits_{j=1}^N \left( \textbf{\emph{x}}_j -a \right)^2 \operatorname*{Var}\left( \bm{\varPhi} _{ij}\right)
$$
$$
=\frac{1}{M} \sum\limits_{j=1}^N \left( \textbf{\emph{x}}_j -a \right)^2
, \eqno(13)
$$
respectively.

According to Theorem 1, we can see that the variance of  $ \textbf{\emph{y}}' $  is related to the constant $ a $. In practice, it is desired that, after gray transformation, the variance of CS samples is as small as possible, i.e.,
 $$
\hat{a}=\operatorname*{argmin}\limits_{{a}} f \left( a \right)=
\frac{1}{M} \sum\limits_{j=1}^N \left( \textbf{\emph{x}}_j -a \right)^2.\eqno(14)
$$

To find an optimal solution, we can take the derivative of the function $ f \left( a \right) $ with respect to $ a $, the outcome is
 $$
\frac{\partial f \left( a \right)} {\partial a} =
\frac{1} {M}\left(2Na-  2 \sum\limits_{j=1}^N \textbf{\emph{x}}_j\right)
 $$
 $$
=\frac{2N} {M}\left(a-  \frac{1} {N} \sum\limits_{j=1}^N \textbf{\emph{x}}_j\right).\eqno(15)
$$

Obviously, when $ a= \frac{1}{N} \sum\limits_{j=1}^N \textbf{\emph{x}}_j $, the derivative of $ f \left( a \right) $ equals zero. It is easy to verify that $ f \left( a \right) $ has minimum value when  $ a= \frac{1}{N} \sum\limits_{j=1}^N \textbf{\emph{x}}_j $. Therefore, we can conclude that the variance of $ \textbf{\emph{y}}' $  is minimum when $ a $  equals the mean value of the original image.

\section{Simulation results}

\begin{figure*}

	\centering
	\begin{minipage}[]{0.189\linewidth}
		\centering
		\centerline{
			\includegraphics[width = \linewidth]{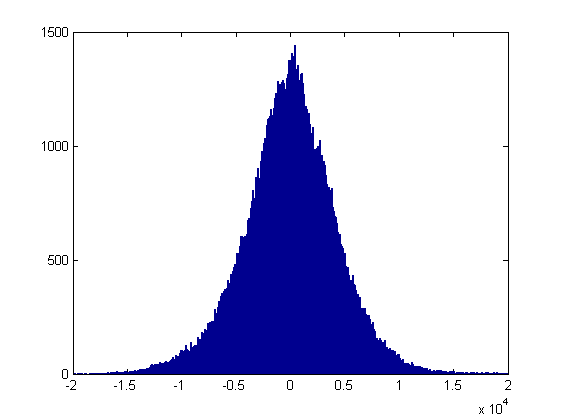}}
		\centerline{(a1)}
		\medskip	
	\end{minipage}
\hspace{0.01 cm}
\begin{minipage}[]{0.189\linewidth}
		\centering
		\centerline{
			\includegraphics[width = \linewidth]{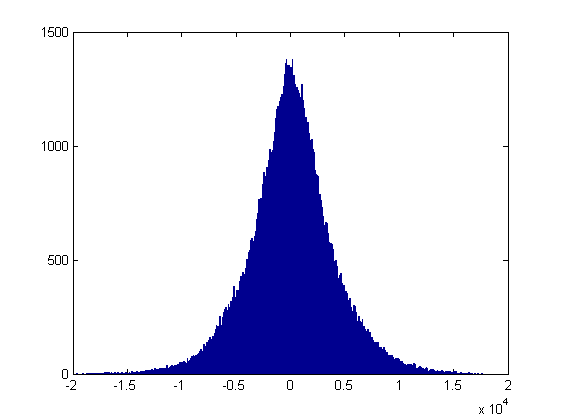}}
		\centerline{(a2)}
		\medskip	
	\end{minipage}
\hspace{0.01 cm}
\begin{minipage}[]{0.189\linewidth}
		\centering
		\centerline{
			\includegraphics[width = \linewidth]{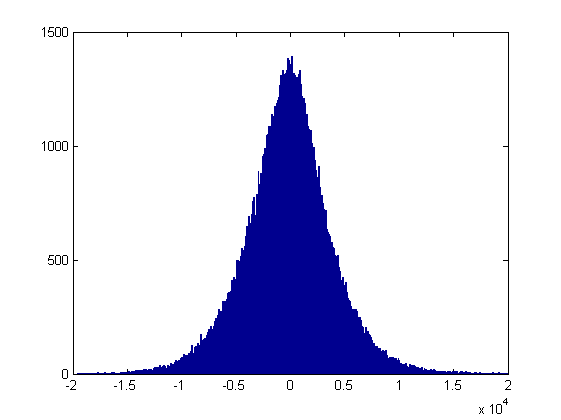}}
		\centerline{(a3)}
		\medskip	
	\end{minipage}
\hspace{0.01 cm}
\begin{minipage}[]{0.189\linewidth}
		\centering
		\centerline{
			\includegraphics[width = \linewidth]{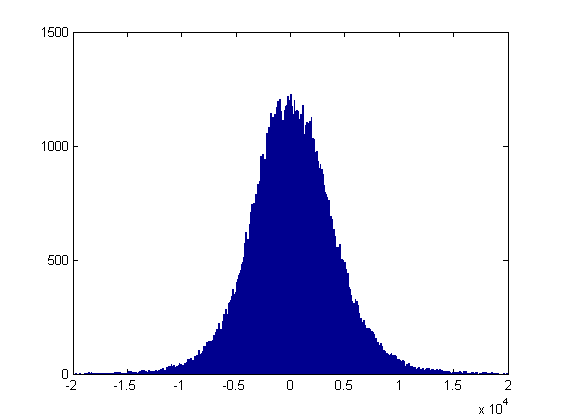}}
		\centerline{(a4)}
		\medskip	
	\end{minipage}
\hspace{0.01 cm}
\begin{minipage}[]{0.189\linewidth}
		\centering
		\centerline{
			\includegraphics[width = \linewidth]{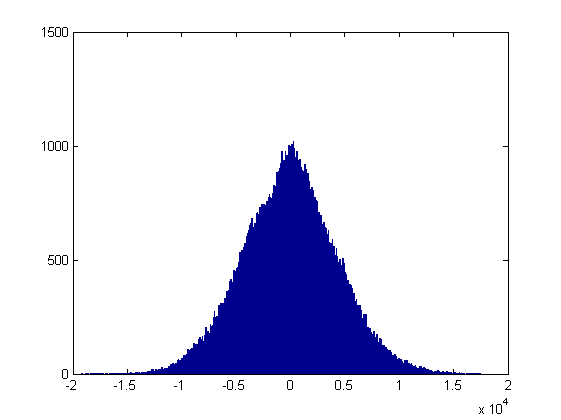}}
		\centerline{(a5)}
		\medskip	
	\end{minipage}
\hspace{0.01 cm}
\begin{minipage}[]{0.189\linewidth}
		\centering
		\centerline{
			\includegraphics[width = \linewidth]{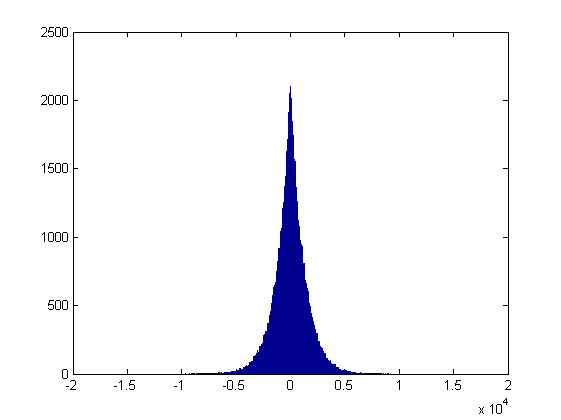}}
		\centerline{(b1)}
		\medskip	
	\end{minipage}
\hspace{0.01 cm}
\begin{minipage}[]{0.189\linewidth}
		\centering
		\centerline{
			\includegraphics[width = \linewidth]{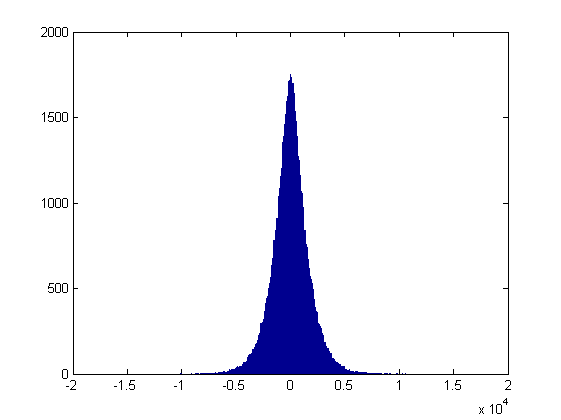}}
		\centerline{(b2)}
		\medskip	
	\end{minipage}
\hspace{0.01 cm}
\begin{minipage}[]{0.189\linewidth}
		\centering
		\centerline{
			\includegraphics[width = \linewidth]{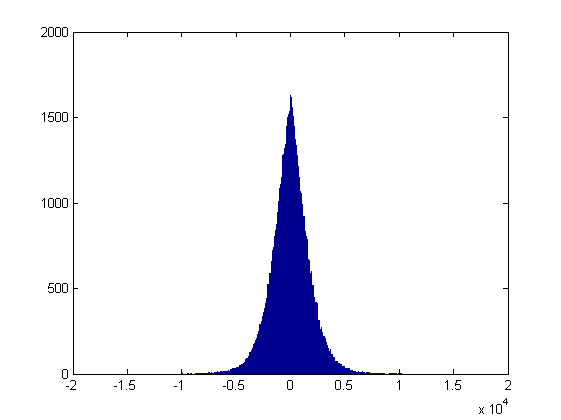}}
		\centerline{(b3)}
		\medskip	
	\end{minipage}
\hspace{0.01 cm}
\begin{minipage}[]{0.189\linewidth}
		\centering
		\centerline{
			\includegraphics[width = \linewidth]{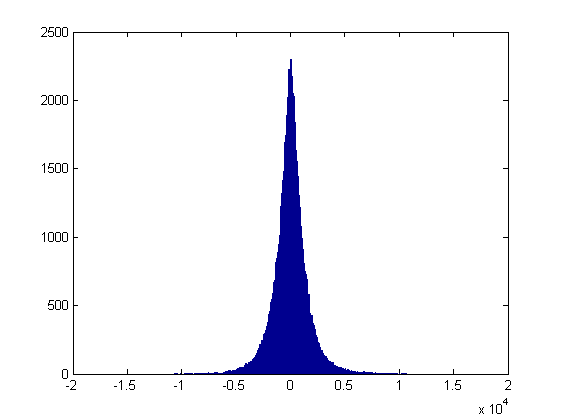}}
		\centerline{(b4)}
		\medskip	
	\end{minipage}
\hspace{0.01 cm}
\begin{minipage}[]{0.189\linewidth}
		\centering
		\centerline{
			\includegraphics[width = \linewidth]{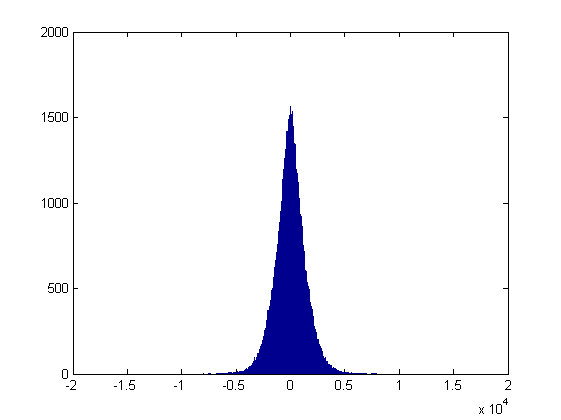}}
		\centerline{(b5)}
		\medskip	
	\end{minipage}
  \caption{Histograms of the CS samples for different schemes. (a1)-(a5) are the histograms of the CS samples for Lena, Peppers, Barbara, Goldhill and Mandrill by using SBCS scheme, respectively; (b1)-(b5) are the histograms of the CS samples for Lena, Peppers, Barbara, Goldhill and Mandrill by using SBCS-GT scheme, respectively.}
\end{figure*}

In this section, a lot of experimental results are reported to verify the performance of the proposed system. Five images, such as Lena, Peppers, Barbara, Goldhill and Mandrill, are used in the experiments. The size of the tested images is $ 512\times 512 $. Gaussian random matrix is applied in the experiments and the sparsity basis is the dualtree discrete wavelet (DDWT) [19]. PSNR is used to evaluate the objective quality of images. To simplify the encoder, we use block CS (BCS) to sample the original image, i.e., the original image is splited into non-overlap blocks firstly and then these blocks are sampled by using CS. The block size is $ 32\times 32 $.

In our proposed method, we use a gray transformation to preprocess the 2D image, which can reduce the variance of CS samples. In the first experiment, we present the probability distribution of CS samples to demonstrate this fact. The sampling rate is 0.5 in this test. Two schemes are considered.

(1) Scrambled BCS (SBCS) [5] scheme. This is a state-of-the-art BCS scheme. Gray transformation is not applied for this scheme.

(2) Scrambled BCS with gray transformation (SBCS-GT) scheme. For this scheme, gray transformation is utilized prior to CS sampling and then the transformed image is sampled by using SBCS.

For both schemes, we use smoothed projected Landweber (SPL) [13] algorithm to recover the image.  In [5], the authors also use a non-local low-rank reconstruction (NLR) method [20] to refine the recovered image. In this letter, we do not refine the reconstructed image by using NLR method for both schemes because the computational complexity of NLR method is very high [21].

The probability distribution of the CS samples for different schemes is shown in Fig. 3. From the figure, we can see that the probability distribution of the CS samples has a large variance and the tail decays slowly for SBCS scheme. Compared with SBCS scheme, SBCS-GT scheme has smaller variance and the tail decays more faster. Therefore, for a given distortion, it is expected that the bit depth required for encoding each CS sample can be reduced considerably by using SBCS-GT scheme.

In our second experiment, the required bit size for encoding each CS sample of the proposed scheme is tested. We quantize the CS sample by using different bit sizes. For the sake of comparison, lossless case is also considered in this test. The PSNR values of the reconstructed images for SBCS scheme and SBCS-GT scheme are shown in Table I and Table II, respectively. From the tables, we can see that 6 bits are enough to encode each CS sample for SBCS-GT scheme. However, for SBCS scheme, 8 bits are required to encode each CS sample. In conclusion, SBCS-GT scheme saves 2 bits for each CS sample in comparison with SBCS scheme.

\begin{table}[]
\setlength{\abovecaptionskip}{0pt}
\setlength{\belowcaptionskip}{2pt}
\caption{The PSNR (in dB) values of the reconstructed images for SBCS scheme with different bit sizes.}
\setlength{\tabcolsep}{1.2mm}
\centering  
\begin{tabular}{lllllll}
\hline
\multicolumn{1}{c}{\multirow{2}{*}{Images}} & \multicolumn{1}{c}{\multirow{2}{*}{Bit sizes}} & \multicolumn{5}{c}{Sampling rate}                                             \\ \cline{3-7}
\multicolumn{1}{c}{}                       & \multicolumn{1}{c}{}                        & 0.1           & 0.2           & 0.3           & 0.4           & 0.5           \\ \hline

\multirow{5}{*}{Lena}         & Lossless case   & 27.64 	& 31.16 	& 33.33 	& 35.01 	& 36.57        \\
                              & 8 bits          & 27.59 	& 31.06 	& 33.13 	& 34.70 	& 36.05        \\
                              & 7 bits          & 27.54 	& 30.87 	& 32.71 	& 34.08 	& 35.09        \\
                              & 6 bits          & 27.45 	& 30.30 	& 31.53 	& 31.65 	& 32.02      \\  \hline

\multirow{5}{*}{Peppers}      & Lossless case   & 28.42 	& 31.83 	& 33.64 	& 35.07 	& 36.32        \\
                              & 8 bits          & 28.32 	& 31.60 	& 33.50 	& 34.79 	& 35.94        \\
                              & 7 bits          & 28.24 	& 31.41 	& 33.21 	& 34.04 	& 35.17       \\
                              & 6 bits          & 27.96 	& 30.74 	& 31.41 	& 31.92 	& 32.20      \\  \hline

\multirow{5}{*}{Barbara}      & Lossless case   & 22.35 	& 23.63 	& 24.89 	& 26.48 	& 28.06        \\
                              & 8 bits          & 22.32 	& 23.58 	& 24.87 	& 26.35 	& 27.97       \\
                              & 7 bits          & 22.09 	& 23.36 	& 24.86 	& 26.26 	& 27.66        \\
                              & 6 bits          & 21.98 	& 23.05 	& 24.47 	& 25.44 	& 26.61      \\  \hline

\multirow{5}{*}{Goldhill}     & Lossless case   & 26.78 	& 28.90 	& 30.30 	& 31.53 	& 32.72        \\
                              & 8 bits          & 26.78 	& 28.83 	& 30.12 	& 31.31 	& 32.40        \\
                              & 7 bits          & 24.39 	& 28.77 	& 30.02 	& 30.80 	& 32.01        \\
                              & 6 bits          & 23.51 	& 28.02 	& 28.99 	& 29.73 	& 30.31      \\  \hline

\multirow{5}{*}{Mandhill}     & Lossless case   & 20.36 	& 21.78 	& 22.84 	& 23.82 	& 24.86        \\
                              & 8 bits          & 20.34 	& 21.78 	& 22.78 	& 23.78 	& 24.78      \\
                              & 7 bits          & 20.18 	& 21.67 	& 22.72 	& 23.75 	& 24.69        \\
                              & 6 bits          & 20.10 	& 21.66 	& 22.63 	& 22.97 	& 24.42      \\  \hline

\end{tabular}
\label{tab1}
\end{table}

\begin{table}[]
\setlength{\abovecaptionskip}{0pt}
\setlength{\belowcaptionskip}{2pt}
\caption{The PSNR (in dB) values of the reconstructed images for SBCS-GT scheme with different bit sizes.}
\setlength{\tabcolsep}{1.2mm}
\centering  
\begin{tabular}{lllllll}
\hline
\multicolumn{1}{c}{\multirow{2}{*}{Images}} & \multicolumn{1}{c}{\multirow{2}{*}{Bit sizes}} & \multicolumn{5}{c}{Sampling rate}                                             \\ \cline{3-7}
\multicolumn{1}{c}{}                       & \multicolumn{1}{c}{}                        & 0.1           & 0.2           & 0.3           & 0.4           & 0.5           \\ \hline

\multirow{5}{*}{Lena}         & Lossless case   & 27.64 	& 31.16 	& 33.33 	& 35.01 	& 36.57        \\
                              & 8 bits          & 27.58 	& 31.15 	& 33.28 	& 34.94 	& 36.43        \\
                              & 7 bits          & 27.56 	& 31.09 	& 33.08 	& 34.78 	& 36.23        \\
                              & 6 bits          & 27.46 	& 30.97 	& 32.76 	& 34.21 	& 35.43      \\  \hline

\multirow{5}{*}{Peppers}      & Lossless case   & 28.42 	& 31.83 	& 33.64 	& 35.07 	& 36.32        \\
                              & 8 bits          & 28.40 	& 31.71 	& 33.63 	& 35.03 	& 36.22        \\
                              & 7 bits          & 28.22 	& 31.68 	& 33.43 	& 34.82 	& 35.94        \\
                              & 6 bits          & 28.11 	& 31.47 	& 33.00 	& 34.04 	& 35.09      \\  \hline

\multirow{5}{*}{Barbara}      & Lossless case   & 22.35 	& 23.63 	& 24.89 	& 26.48 	& 28.06        \\
                              & 8 bits          & 22.31 	& 23.63 	& 24.88 	& 26.41 	& 27.97        \\
                              & 7 bits          & 22.21 	& 23.57 	& 24.86 	& 26.25 	& 27.93        \\
                              & 6 bits          & 22.11 	& 23.48 	& 24.79 	& 26.16 	& 27.81      \\  \hline

\multirow{5}{*}{Goldhill}     & Lossless case   & 26.78 	& 28.90 	& 30.30 	& 31.53 	& 32.72        \\
                              & 8 bits          & 26.64 	& 28.89 	& 30.28 	& 31.47 	& 32.65        \\
                              & 7 bits          & 24.87 	& 28.77 	& 30.22 	& 31.40 	& 32.48        \\
                              & 6 bits          & 24.29 	& 28.63 	& 29.80 	& 30.87 	& 31.91      \\  \hline

\multirow{5}{*}{Mandhill}     & Lossless case   & 20.36 	& 21.78 	& 22.84 	& 23.82 	& 24.86        \\
                              & 8 bits          & 20.27 	& 21.73 	& 22.80 	& 23.72 	& 24.83       \\
                              & 7 bits          & 20.17 	& 21.72 	& 22.79 	& 23.42 	& 24.78        \\
                              & 6 bits          & 20.09 	& 21.70 	& 22.75 	& 23.23 	& 24.76      \\  \hline

\end{tabular}
\label{tab1}
\end{table}

\begin{table}[]
\setlength{\abovecaptionskip}{0pt}
\setlength{\belowcaptionskip}{2pt}
\caption{Comparison of the compression performance for different schemes.}
\setlength{\tabcolsep}{1.2mm}
\centering  
\begin{tabular}{lllllll}
\hline
\multirow{2}{*}{Images}                    & \multirow{2}{*}{Schemes} & \multicolumn{5}{c}{Bit rate} \\ \cline{3-7}
                                        &                         & 0.2	    & 0.4	& 0.6   & 0.8	  & 1.0 \\ \hline
\multicolumn{1}{c}{\multirow{2}{*}{Lena}} & SBCS                  & 20.68	& 26.75	& 28.41 & 29.33	& 30.23   \\
\multicolumn{1}{c}{}                    & SBCS-GT                 & 24.26	& 27.21	& 28.95	& 30.42	&  31.46 \\ \hline
\multicolumn{1}{c}{\multirow{2}{*}{Peppers}} & SBCS               &  21.38	& 26.71	& 28.96	& 29.84	& 30.69   \\
\multicolumn{1}{c}{}                    & SBCS-GT                 & 22.86	& 27.55	& 29.54	& 30.82	& 31.66 \\ \hline
\multicolumn{1}{c}{\multirow{2}{*}{Barbara}} & SBCS               & 20.18	& 21.77	& 22.39	& 23.08	& 23.62   \\
\multicolumn{1}{c}{}                    & SBCS-GT                 & 20.53    & 22.18	& 22.64	& 23.34	& 23.98 \\ \hline
\multicolumn{1}{c}{\multirow{2}{*}{Goldhill}} & SBCS              & 19.95	& 21.07	& 27.02	& 27.72	& 28.19   \\
\multicolumn{1}{c}{}                    & SBCS-GT                 & 20.81	& 22.10	& 27.73	& 28.57	& 29.15 \\ \hline
\multicolumn{1}{c}{\multirow{2}{*}{Mandrill}} & SBCS              & 19.45	& 19.95	& 20.60	& 21.22	& 21.67   \\
\multicolumn{1}{c}{}                    & SBCS-GT                 & 20.06	& 20.17	& 20.90	& 21.43	& 22.01 \\ \hline
\multicolumn{1}{c}{\multirow{2}{*}{Average}} & SBCS               & 20.33	& 23.25	& 25.48	& 26.24	& 26.88  \\
\multicolumn{1}{c}{}                    & SBCS-GT                 & 21.70	 & 23.84	 & 25.95	 & 26.92	 & 27.65 \\ \hline
\end{tabular}
\end{table}

In the last experiment, we compare the compression performance between SBCS scheme and SBCS-GT scheme. For both schemes, the bit depth for each CS sample is 6 bits. We alter the sampling rate to control the compression ratio. The compression performance for the two schemes is shown in Table III. It can seen from the table that SBCS-GT scheme outperforms the traditional SBCS scheme in terms of compression performance. In conclusion, since gray transformation can reduce the bit depth required for each CS sample significantly, the proposed method can considerably improve the compression performance of CS-based image coding.

\section{Conclusions}
In this letter, a novel CS-based image coding system by using gray transformation is proposed. Firstly, we use a gray transformation to preprocess the original image. Then, we apply CS to sample the transformed image. Since gray transformation can reduce the bit depth for encoding each CS sample significantly, the proposed method can substantially improve the compression performance of CS-based image coding. The effectiveness of the proposed method is verified by simulation results.


%

\ifCLASSOPTIONcaptionsoff
  \newpage
\fi


\begin{thebibliography}{1}


\bibitem{IEEEhowto:kopka}

D. L. Donoho, ``Compressed sensing,'' \emph{IEEE Trans. on Information Theory}, vol. 52, no. 4, pp. 1289-1306, 2006.

\bibitem{IEEEhowto:kopka}

R. G. Baraniuk, ``Compressive sensing,'' \emph{IEEE Signal Processing Magazine}, vol. 24, no. 4, pp. 118-121, 2007.

\bibitem{IEEEhowto:kopka}

B. Zhang, D. Xiao, and Y. Xiang, ``Robust coding of encrypted images via 2D compressed sensing,'' \emph{IEEE Trans. on Multimedia}, 2020, Early Access.


\bibitem{IEEEhowto:kopka}

V. Pudi, A. Chattopadhyay, and K. Y. Lam, ``Efficient and lightweight quantized compressive sensing using $\mu$-Law,'' \emph{in Proc. of 2018 IEEE International Symposium on Circuits and Systems (ISCAS)}, pp. 1-4, 2018.

\bibitem{IEEEhowto:kopka}

Z. Chen, X. Hou, X. Qian, and G. Chen, ``Efficient and robust image coding and transmission based on scrambled block compressive sensing,'' \emph{IEEE Trans. on Multimedia}, vol. 20, no. 7, pp. 1610-1621, 2018.

\bibitem{IEEEhowto:kopka}

Z. Chen, X. Hou, X. Shao, C. Gong, X. Qian, Y. Huang, and S. Wang, ``Compressive Sensing multi-layer residual coefficients for image coding,'' \emph{IEEE Trans. on Circuits and Systems for Video Technology}, vol. 30, no. 4, pp. 1109-1120, 2020.


\bibitem{IEEEhowto:kopka}
Y. Rivenson and A. Stern, ``Compressed imaging with a separable sensing operator,'' \emph{IEEE Signal Processing Letters}, vol. 16, no. 6, pp. 449-452, 2009.


\bibitem{IEEEhowto:kopka}

B. Zhang, Y. Liu, J. Zhuang, K. Wang, and Y. Cao, ``Matrix permutation meets block compressed sensing,'' \emph{Journal of Visual Communication and Image Representation}, vol. 60, pp. 69-78, 2019.

\bibitem{IEEEhowto:kopka}

L. Wang, X. Wu, and G. Shi, ``Binned progressive quantization for compressive sensing,'' \emph{IEEE Trans. on Image Processing}, vol. 21, no. 6, pp. 2980-2990, 2012.

\bibitem{IEEEhowto:kopka}

A. S. Unde and P. P. Deepthi, ``Rate-distortion analysis of structured sensing matrices for block compressive sensing of images,'' \emph{Signal Processing: Image Communication}, vol. 65, pp. 115-127, 2018.


\bibitem{IEEEhowto:kopka}
C. Zhao, J. Zhang, S. Ma, and W. Gao, ``Compressive-sensed image coding via stripe-based DPCM,'' \emph{in Proc. of Data Compression Conference}, pp. 171-180, 2016.

\bibitem{IEEEhowto:kopka}
S. Mun and J. E. Fowler, ``DPCM for quantized block-based compressed sensing of images,'' \emph{in Proc. of European Signal Processing Conference (EUSIPCO)}, pp. 1424-1428, 2012.


\bibitem{IEEEhowto:kopka}

S. Mun and J. E. Fowler, ``Block compressed sensing of images using directional transforms,'' \emph{in Proc. of International Conference on Image Processing}, pp. 3021-3024, 2009.

\bibitem{IEEEhowto:kopka}

X. Yuan, ``Generalized alternating projection based total variation minimization for compressive sensing,'' \emph{in Proc. of International Conference on Image Processing}, pp. 2539-2543, 2016.


\bibitem{IEEEhowto:kopka}

W. Dong, G. Shi, X. Li, Y. Ma, and F. Huang, ``Compressive sensing via nonlocal low-rank regularization,'' \emph{IEEE Trans. on Image Processing}, vol. 23, no. 8, pp. 3618-3632, 2014.

\bibitem{IEEEhowto:kopka}

S. Chen, M. A. Saunders, and D. L. Donoho, ``Atomic decomposition by basis pursuit,'' \emph{SIAM Review}, vol. 43, no. 1, pp. 129-159, 2001.

\bibitem{IEEEhowto:kopka}

J. Tropp and A. Gilbert, ``Signal recovery from random measurements via orthogonal matching pursuit,'' \emph{IEEE Trans. on Information Theory}, vol. 53, no. 12, pp. 4655-4666, 2007.

\bibitem{IEEEhowto:kopka}

D. Needell and J. A. Tropp, ``Cosamp: iterative signal recovery from incomplete and inaccurate samples,'' \emph{Applied and Computational Harmonic Analysis}, vol. 26, no. 3, pp. 301-321, 2008.


\bibitem{IEEEhowto:kopka}
J. Yang, Y. Wang, W. Xu, and Q. Dai, ``Image coding using dual-tree discrete wavelet transform,'' \emph{IEEE Trans. on Image Processing}, vol. 17, no. 9, pp. 1555-1569, 2008.

\bibitem{IEEEhowto:kopka}
W. Dong, G. Shi, X. Li, Y. Ma, and F. Huang, ``Compressive sensing via nonlocal low-rank regularization,'' \emph{IEEE Trans. on Image Processing}, vol. 23, no. 8, pp. 3618-3632, 2014.


\bibitem{IEEEhowto:kopka}
 X. Yuan and R. Haimi-Cohen, ``Image compression based on compressive sensing: end-to-end comparison with JPEG,'' \emph{IEEE Transactions on Multimedia}, 2020, Early Access.




\end{thebibliography}
\end{document}